\documentclass[aps,pra,twocolumn,superscriptaddress]{revtex4-1}
\usepackage{graphicx}
\usepackage{amsmath}
\usepackage{color}

\begin{document}


\title{Imaging of Relaxation Times and Microwave Field Strength \\ in a Microfabricated Vapor Cell}


\author{Andrew Horsley}
\email[]{andrew.horsley@unibas.ch}
\author{Guan-Xiang Du}
\affiliation{Departement Physik, Universit\"{a}t Basel, Switzerland}
\author{Matthieu Pellaton}
\author{Christoph Affolderbach}
\author{Gaetano Mileti}
\affiliation{Laboratoire Temps-Fr\'{e}quence, Institut de Physique, Universit\'{e} de Neuch\^{a}tel, Switzerland}
\author{Philipp Treutlein}
\email[]{philipp.treutlein@unibas.ch}
\affiliation{Departement Physik, Universit\"{a}t Basel, Switzerland}



\date{\today}

\begin{abstract}
We present a new characterisation technique for atomic vapor cells, combining time-domain measurements with absorption imaging to obtain spatially resolved information on decay times, atomic diffusion and coherent dynamics. The technique is used to characterise a 5~mm diameter, 2~mm thick microfabricated Rb vapor cell, with N$_2$ buffer gas, placed inside a microwave cavity. Time-domain Franzen and Ramsey measurements are used to produce high-resolution images of the population ($T_1$) and coherence ($T_2$) lifetimes in the  cell, while Rabi measurements yield images of the $\sigma_-$, $\pi$ and $\sigma_+$ components of the applied microwave magnetic field. For a cell temperature of 90$^{\circ}$C, the $T_1$ times across the cell centre are found to be a roughly uniform $265\,\mu$s, while the $T_2$ times peak at around $350\,\mu$s. We observe a `skin' of reduced $T_1$ and $T_2$ times around the edge of the cell due to the depolarisation of Rb after collisions with the silicon cell walls. Our observations suggest that these collisions are far from being 100$\%$ depolarising, consistent with earlier observations made with Na and glass walls. Images of the microwave magnetic field reveal regions of optimal field homogeneity, and thus coherence. Our technique is useful for vapor cell characterisation in atomic clocks, atomic sensors, and quantum information experiments.
\end{abstract}


\maketitle

\section{\label{sec:introduction} Introduction}

The use of alkali vapor cells in atomic physics has a history extending back several decades~\cite{Franzen1959,Arditi1964}, and has led to important applications in precision measurement~\cite{Budker2007,Micalizio2012} and quantum information~\cite{Julsgaard2001}. Recent years have seen great interest in newly developed miniaturised and microfabricated vapor cells, with sizes on the order of a few millimeters or smaller. Applications include miniaturised atomic clocks~\cite{Knappe2004,Pellaton2012}, gyroscopes~\cite{Donley2009}, and magnetometers measuring both DC~\cite{Balabas2006,Shah2007,Schwindt2007,Scholtes2012} and radio-frequency~\cite{Savukov2005} fields. As new applications, one of our groups has recently demonstrated imaging of microwave magnetic fields using a vapor cell~\cite{Bohi2012,Bohi2010}, and detection of microwave electric fields has been reported in Ref.~\cite{Sedlacek2012}. Thanks to microfabrication, vapor cells have been miniaturised to a point where spatially resolved information on their properties, and on the external fields applied to them, is essential to their characterisation and performance.

In this paper, we describe a new characterisation technique, applying time-domain Franzen~\cite{Franzen1959}, Ramsey~\cite{Ramsey1956}, and Rabi~\cite{Gentile1989} measurements and absorption imaging~\cite{Ketterle1999} to a microcell. Time-domain measurements in vapor cells are currently experiencing a renaissance in interest~\cite{Micalizio2012b}. Absorption imaging is well established in use with ultracold atoms~\cite{Ketterle1999}, providing single-atom sensitivity~\cite{Streed2012}, and micrometer spatial resolution~\cite{Esteve2008}, however its use with room-temperature atoms is a relatively unexplored area. We use these tools to characterise a microfabricated vapor cell~\cite{DiFrancesco2010,Pellaton2012} and a microwave cavity designed for compact vapor cell atomic clocks~\cite{Mileti1992}, obtaining spatially resolved images of decay times in the cell and images of the microwave field applied to the cell.

This paper is organised as follows. In section~\ref{sec:experimental_setup}, we describe the experimental setup and features of our vapor cell. In section~\ref{sec:PD_measurements} we introduce the Franzen, Ramsey, and Rabi experimental sequences, and some basic measurements using a photodiode for detection. We begin section~\ref{sec:imaging} by describing our adaptation of absorption imaging to vapor cells. We then present images of the $T_1$ and $T_2$ times, and of the atomic populations in the optically pumped steady state. We investigate Rb-wall collisions and describe the $T_1$ relaxation by modelling optical pumping, diffusion and collisional relaxation in the cell, and finish section~\ref{sec:imaging} with polarisation-resolved images of the microwave magnetic field amplitude. We conclude, and discuss future directions, in section~\ref{sec:outlook}.

\section{\label{sec:experimental_setup} Experimental Setup and Initial Characterisation}

\begin{figure}
\includegraphics[width=0.5\textwidth]{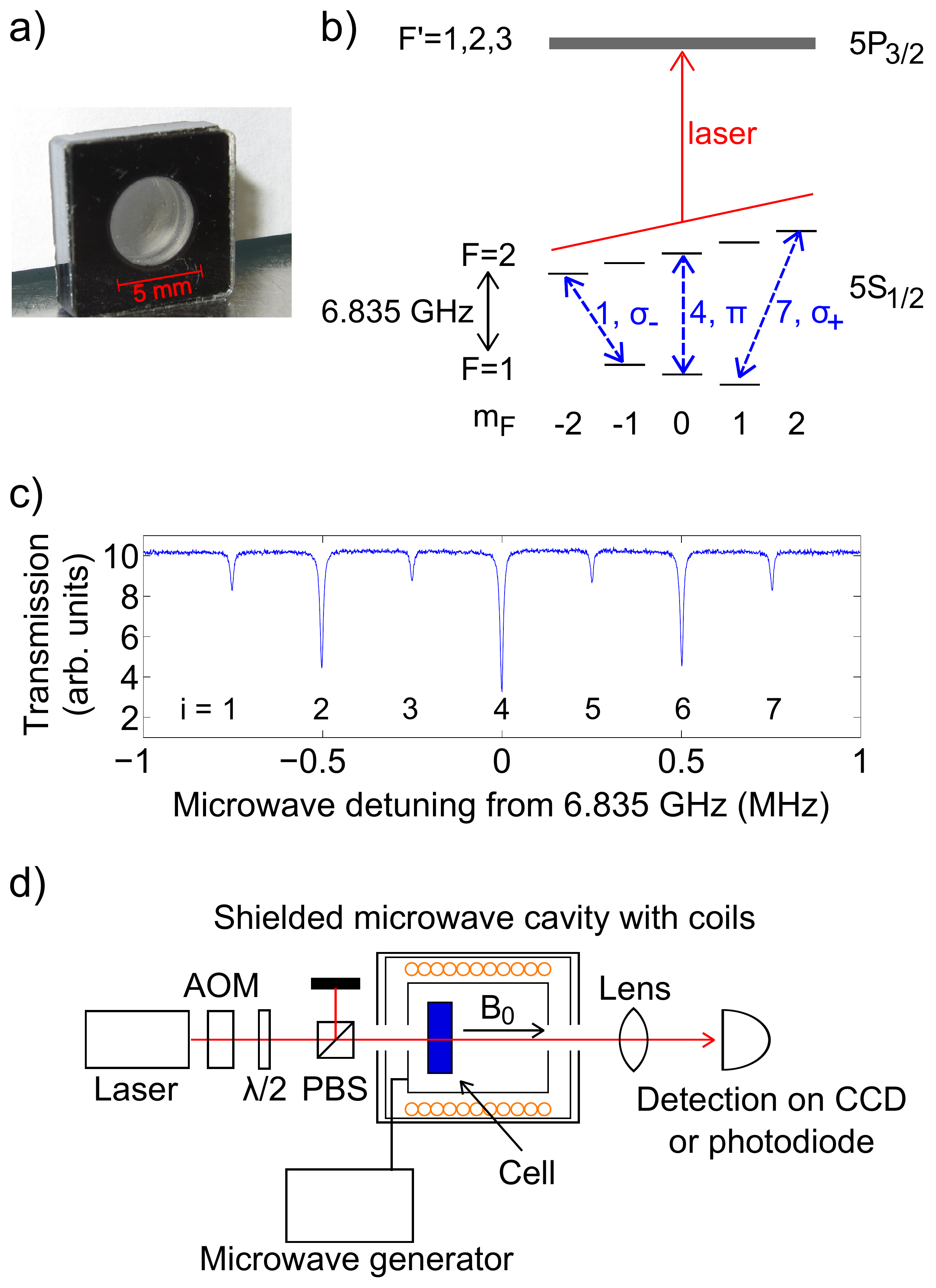}%
\caption{\label{fig:experiment_setup}a) The microfabricated vapor cell used in this paper, with glass windows and a silicon frame; b) The $^{87}$Rb D2 line. Due to Doppler and collisional broadening on the optical transitions, the excited state hyperfine levels $F'$ are not resolved. Transitions between the Zeeman-split m$_F$ levels of the ground state hyperfine structure can be individually addressed by the microwave field. The three hyperfine transitions used in this work ($i=1,4,7$) are shown in dotted blue; c) A double resonance spectrum, showing laser transmission through the cell as the microwave frequency is scanned. Transmission is reduced whenever the microwave comes on resonance with a hyperfine transition; d) The experimental setup.}
\end{figure}

\subsection{Equipment and Setup}

We use the microfabricated cell shown in Figure~\ref{fig:experiment_setup}a. The cell has a $5\,\mathrm{mm}\times2\,\mathrm{mm}$ internal diameter and thickness, and contains natural abundance Rb and $63\pm2\,\mathrm{mbar}$ of N$_2$ buffer gas~\cite{Pellaton2012}. This buffer gas pressure was measured at 80$^{\circ}$C from the line-shift induced on the $^{87}$Rb clock transition \cite{Pellaton2012}, using the coefficients provided in~\cite{Vanier1982}. The cell is inserted into a microwave cavity~\cite{Mileti1992}, which is tuned to have its resonance frequency at the 6.835~GHz ground-state hyperfine splitting of $^{87}$Rb. The cavity is surrounded by a solenoid coil that provides a static magnetic field of 35~$\mu$T, parallel to the direction of laser propagation (see Figure~\ref{fig:experiment_setup}d). The resulting 0.25~MHz Zeeman splitting between transitions allows all seven $^{87}$Rb hyperfine transitions to be individually addressed, as shown in the double-resonance spectrum of Figure~\ref{fig:experiment_setup}c. A temperature control system is used to heat the cell and actively stabilise its temperature to within a few parts in $10^4$, and an outer double-layer of $\mu$-metal provides magnetic shielding. Except when otherwise noted, the cell temperature was set to 90$^{\circ}$C for all data presented in this paper.

We use a grating stabilised diode laser emitting linearly polarised light at 780~nm, frequency stabilised using saturated absorption spectroscopy to the $F=2 \rightarrow F'=2,3$ crossover peak of the $^{87}$Rb D2 line (5S$_{1/2}\rightarrow5$P$_{3/2}$). Doppler and collisional broadening ensure that the $F=2$ ground state is coupled to all of the $F'=1,2,3$ excited state hyperfine levels (see Figure~\ref{fig:experiment_setup}b). An acousto-optical modulator (AOM), driven at 80~MHz, is used to provide switching with a rise time below 100~ns. A single laser beam is used for both optical pumping~\cite{Happer1972} and absorption measurements on the atoms. Microwave signals near 6.835~GHz are produced by a frequency generator (HP8304B), and passed through a switch and an amplifier before being coupled into the cavity.

\subsection{Hyperfine (Microwave) Transitions}

There are nine possible hyperfine transitions between the $^{87}$Rb ground states, shown in Figure~\ref{fig:experiment_setup}b, three from each $m_F$ level of $F=1$. 
Two degenerate pairs of transitions leave us with seven resonances, which we label $i=1\ldots7$, in order of increasing frequency. We address three (non-degenerate) hyperfine transitions in this work: $i=1$, 4, and 7, or, using $|F,m_F\rangle$ notation: $|1,-1\rangle\rightarrow|2,-2\rangle$, $|1,0\rangle\rightarrow|2,0\rangle$, and $|1,+1\rangle\rightarrow|2,+2\rangle$. These are transitions corresponding to $\sigma_-$, $\pi$, and $\sigma_+$ polarization components of the microwave magnetic field, respectively. $i=4$ represents the `clock transition', exploited in atomic clocks~\cite{Camparo2007}.

The hyperfine transitions are shown in Figure~\ref{fig:experiment_setup}c as a double-resonance spectrum~\cite{Camparo2007}. The spectrum is produced by scanning the frequency of the microwave as the laser illuminates the cell. For this measurement, both the microwave and laser are continuously on. Whenever the microwave comes onto resonance with a hyperfine transition, the optically pumped $F=2$ state is repopulated. This results in a dip in the transmission of the laser, which is recorded by a photodiode. The $\pi$-transitions in Figure~\ref{fig:experiment_setup}c, $i=2,4,6$, are the strongest, as the microwave cavity is designed to operate in a mode where the $\pi$-component dominates.

\subsection{Experiment Sequences}

In this paper we mostly use pulsed experiments to characterize the vapor cell.
In a typical sequence (see section~\ref{sec:PD_measurements}), we first apply an optical pumping pulse to the vapor that depopulates the $F=2$ state. It is followed by microwave pulses that coherently manipulate the atomic hyperfine state. Finally, we measure the optical density (OD) in the $F=2$ state with a probe pulse of the same frequency and intensity, but much shorter duration than the optical pumping pulse, in order to minimise optical pumping during the probe pulse. 
For incident and transmitted probe intensities of $I_0$ and $I_t$, respectively, the OD is defined as
\begin{equation}
\mathrm{OD}=-\ln(I_t/I_0).
\end{equation}
Detection is performed using either a photodiode (Thorlabs DET10A/M), or absorption imaging on a CCD camera (Guppy Pro F031B). Details on the two detection methods are given in sections~\ref{sec:PD_measurements} and \ref{sec:imaging}, respectively.

\subsection{Optical Density as a Function of Temperature}

\begin{figure}
\includegraphics[width=0.4\textwidth]{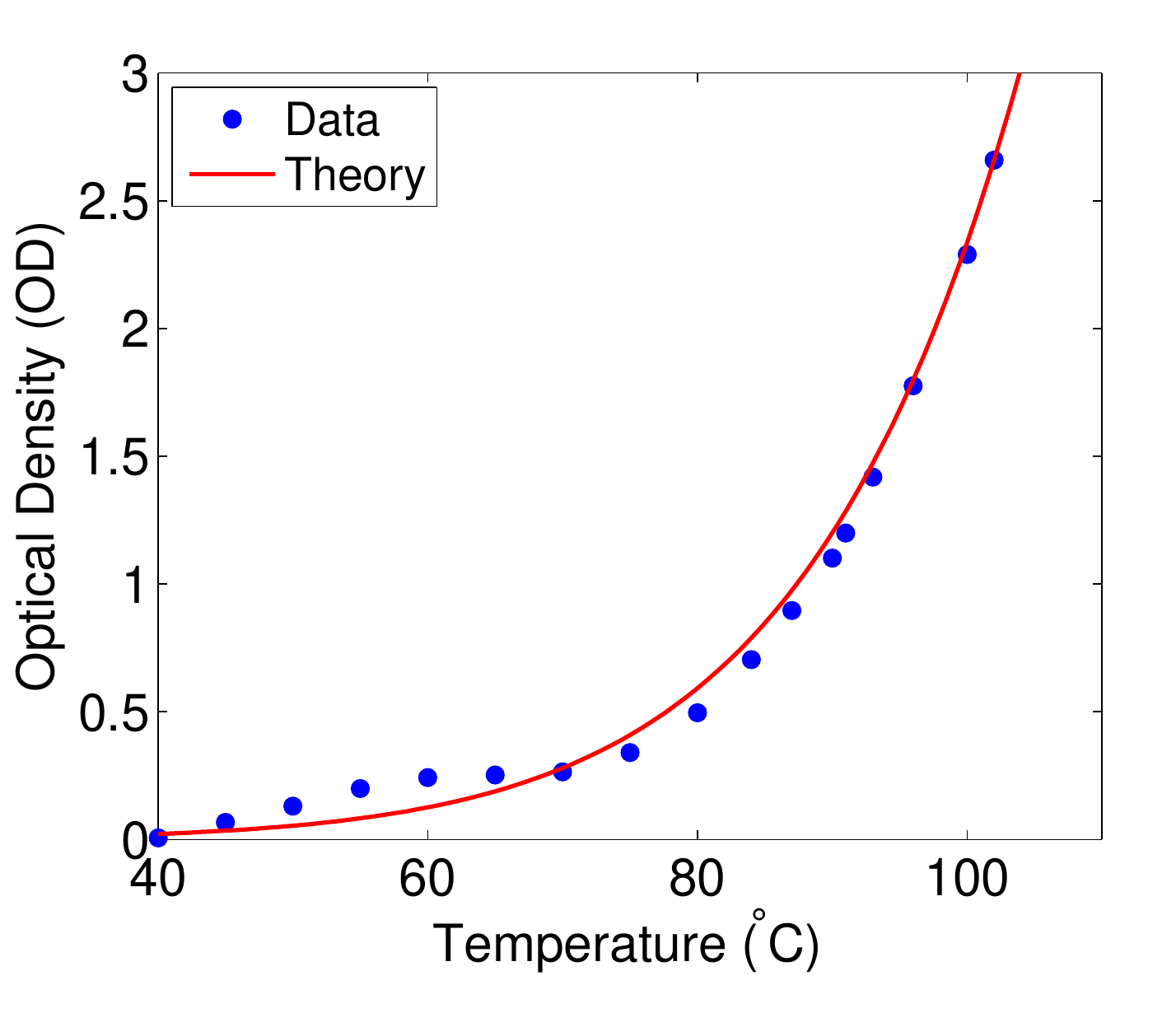}%
\caption{\label{fig:ODvsT} Optical density of the cell as a function of temperature. The theory curve has been produced using the model of Ref.~\cite{Siddons2008}, modified to include pressure and Rb dipole-dipole broadening. The theory has no free parameters.}
\end{figure}

The OD of the vapor in the cell is shown as a function of temperature in Figure~\ref{fig:ODvsT}. Transmission through the centre of the cell of a 2~mm diameter, low intensity ($I_0 <600\,\mu\mathrm{W}/\mathrm{cm}^2$) laser beam was measured with a photodiode. In this case, no optical pumping or microwave pulses were applied. The model described in Ref.~\cite{Siddons2008}, modified to include pressure broadening due to the buffer gas as in Ref.~\cite{Rotondaro1997,Weller2012} and broadening due to Rb dipole-dipole interactions~\cite{Weller2011}, is compared to the data. The agreement is good considering that the theory has no free parameters.

\section{\label{sec:PD_measurements} Time Domain Measurements Without Spatial Resolution}

We use three sequence types in this work: Franzen~\cite{Franzen1959}, Ramsey~\cite{Ramsey1956}, and Rabi~\cite{Gentile1989}. Franzen, or relaxation-in-the-dark, sequences are all-optical, and are used to obtain $T_1$ times. Ramsey sequences provide both $T_1$ and $T_2$ times. The $T_1$ times refer to population relaxation between all $F=1$ and $F=2$ sublevels, whilst the $T_2$ times are specific for the particular hyperfine $m_F$ transition probed. Rabi sequences provide information about the microwave magnetic fields strengths applied to the cell.

We performed a first characterisation of the cell using a photodiode as the detector. When using the photodiode, the transmission of the probe laser pulse is measured 10~$\mu$s after its start, in order to accommodate the photodiode response time. A laser intensity of $\approx 5\,\mathrm{mW}/\mathrm{cm}^2$ was used in  the measurements described in this section, with the beam partially covering the cell. Scanning the laser intensity from $0.1\,\mathrm{mW}/\mathrm{cm}^2$ to $10\,\mathrm{mW}/\mathrm{cm}^2$  produced no apparent variation in relaxation times. This indicates that the small, constant amount of optical pumping induced by the first 10~$\mu$s of the probe pulse does not greatly affect the measured time constants. Unless otherwise stated, uncertainties are taken from the 68$\%$ confidence bounds of fitting to the data.

\begin{figure}
\includegraphics[width=0.5\textwidth]{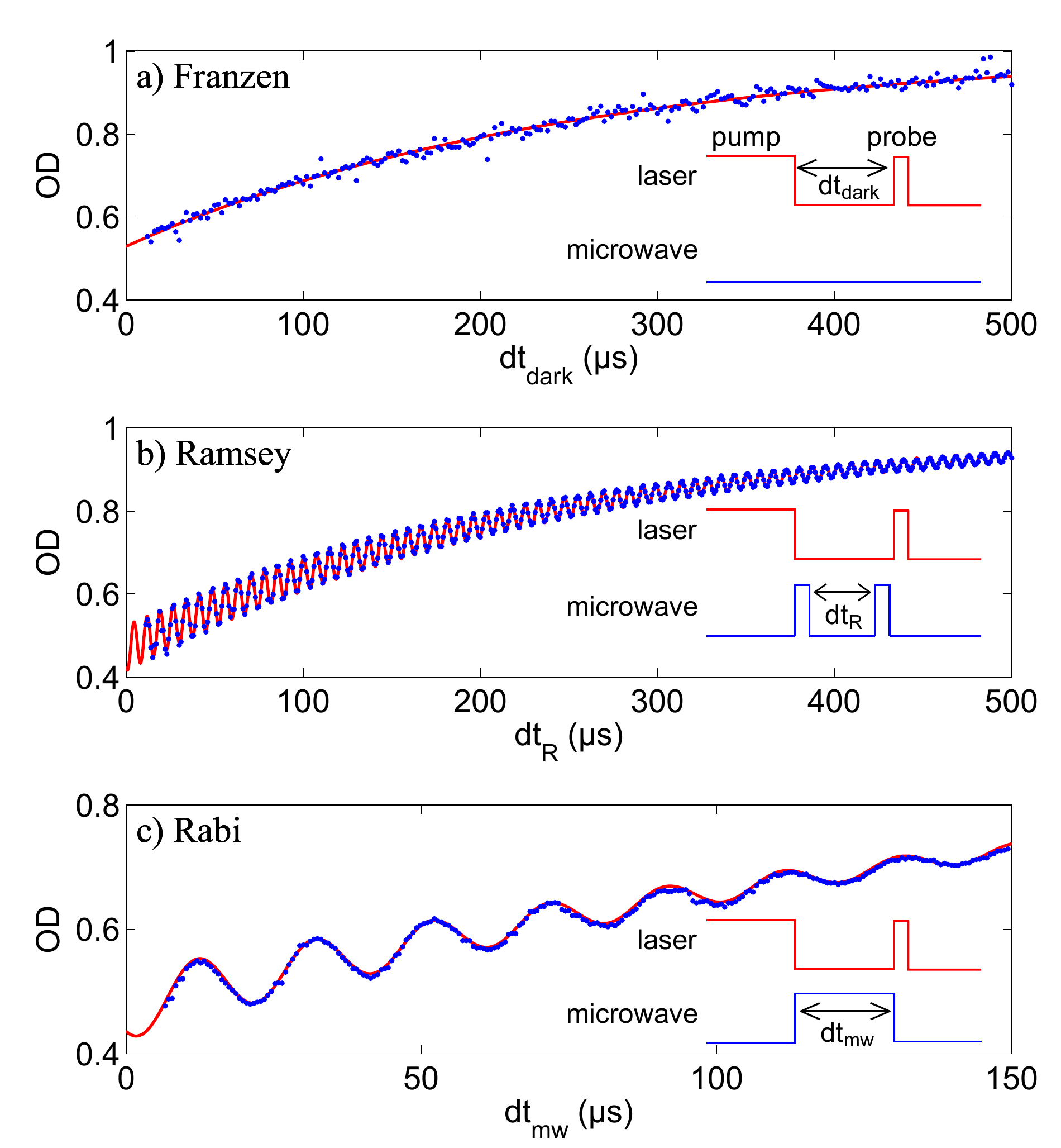}%
\caption{\label{fig:franzen_ramsey_rabi_inset} Cell OD response to a) Franzen, b) Ramsey, and c) Rabi sequences, recorded using a photodiode. Data is shown as blue dots, while the fitting curves (described in the text) are in red. Note the different scale in (c). The insets show the laser and microwave sequences used. The OD increases with laser dark time, as the hyperfine population difference relaxes.}
\end{figure}

\subsection{\label{sec:PD_franzen} Franzen Measurements}

We begin a Franzen sequence by optical hyperfine pumping of the atoms for some milliseconds, depopulating the $F=2$ ground state and reducing the OD of the cell~\cite{Happer1972}. The laser beam is then switched off with the AOM, and the pumped population difference relaxes at a rate $1/T_1$. After a time $dt_{dark}$, we measure the OD with the probe pulse. Scanning $dt_{dark}$ allows us to observe the hyperfine population relaxation and to determine $T_1$.

Figure~\ref{fig:franzen_ramsey_rabi_inset}a shows data from an example Franzen sequence. We fit the following equation to the data:
\begin{equation}\label{eq:franzen_fit}
\mathrm{OD} = A - B\exp(-dt_{dark}/T_1),
\end{equation}
where $A$, $B$, and $T_1$ are fitting parameters. This yields $T_1=(244\pm 6)$~$\mu$s. We neglect possible tensorial hyperfine relaxation, with different relaxation rates for different $m_F$ states, as we observed no significant variation in $T_1$ when scanning the laser polarisation (which scans the relative population of $m_F$ states after optical pumping). The simple nature of the Franzen data and the fitting equation results in fast fitting and robust $T_1$ values.

\subsection{\label{sec:PD_ramsey} Ramsey Measurements}

In Ramsey sequences~\cite{Ramsey1956}, we introduce two microwave pulses between the pump and probe laser pulses of the Franzen sequence. The first pulse creates a coherent superposition of the two hyperfine $m_F$ states that are coupled by the microwave. During the subsequent free evolution of duration $dt_R$, the atomic superposition state accumulates a phase relative to the microwave local oscillator. The second microwave pulse converts this phase into a population difference between the hyperfine states. By scanning $dt_R$, oscillations of the atomic population are recorded. Each microwave pulse is nominally a $\pi/2$ pulse, however variation in the microwave field across the cell (see section~\ref{sec:imaging}) results in atoms experiencing a range of pulse areas. For a given microwave power setting, the nominal $\pi/2$ pulse length is obtained by performing a Rabi sequence using a broad laser beam that illuminates the entire cell, and measuring the Rabi oscillation period on a photodiode. The $\pi/2$ length is then 1/4 of this period. Ramsey sequences are robust to laser and microwave field induced decoherence, as the majority of the atomic evolution occurs in the dark, with the microwave and optical fields off. As such, they provide a good measure of the $T_2$ time of the cell.

Figure~\ref{fig:franzen_ramsey_rabi_inset}b shows an example Ramsey sequence. The microwave power at the input to the cavity was 29.8~dBm. To record Ramsey oscillations in time, the microwave was slightly detuned by $\delta$ from the $i=4$ transition. Although the data is only shown up to 500~$\mu$s, Ramsey oscillations are still clearly visible at evolution times past 1.2~ms. The data is fit with the equation
\begin{eqnarray}\label{eq:ramsey_fit}
\mathrm{OD} &=& A - B\exp(-dt_R/T_1) \nonumber \\
& &+ C\exp(-dt_R/T_2)\sin(\delta \, dt_R + \phi)
\end{eqnarray}
Where $A$, $B$, $C$, $\phi$, $T_1$, $T_2$, and $\delta$ are fitting parameters. The fit gives the two relaxation times as $T_1=(245\pm 0.5)\,\mu$s and $T_2=(322\pm 4)\,\mu$s. The $T_1$ time is in excellent agreement with that obtained from the Franzen measurement. The exact detuning of the microwave from resonance is given by the Ramsey oscillation frequency, $\delta=2\pi \times (135.764\pm 0.006)$~kHz.
The measured $T_2$ is specific to the clock transition. Tuning the microwave to  field-sensitive transitions ($i\neq4$ in Figure~\ref{fig:experiment_setup}), we see  $T_2$  drop by a factor of two to three. This is primarily due to dephasing introduced by inhomogeneities in the static magnetic field.

\subsection{\label{sec:PD_rabi} Rabi Measurements}

A Rabi sequence consists of a single microwave pulse applied during the dark time between the laser pumping and probe pulses~\cite{Gentile1989}. The microwave pulse drives Rabi oscillations between the two resonantly coupled $m_F$ sublevels of $F=1$ and $F=2$, at a frequency proportional to the microwave magnetic field strength. This allows us to use Rabi sequences to measure each vector component of the microwave magnetic field~\cite{Bohi2012,Bohi2010}. By tuning the microwave frequency to transitions $i=1$, 4, and 7, we are sensitive to the $\sigma_-$, $\pi$, and $\sigma_+$ components of the microwave magnetic field, respectively. The magnitude of the microwave magnetic field components is obtained using the equations \cite{Bohi2012}
\begin{eqnarray}
\nonumber B_{-} = \frac{1}{\sqrt{3}}\frac{\hbar}{\mu_B}\Omega_1, \\
\label{eq:mw_Bfield} B_{\pi} = \frac{\hbar}{\mu_B}\Omega_4, \\
\nonumber B_{+} = \frac{1}{\sqrt{3}}\frac{\hbar}{\mu_B}\Omega_7, \nonumber
\end{eqnarray}
where $\Omega_i$ is the Rabi frequency for oscillations on transition $i$.

An example Rabi sequence is shown in Figure~\ref{fig:franzen_ramsey_rabi_inset}c. The microwave power at the input to the cavity was 27.8~dBm, and the microwave frequency was tuned exactly to the $i=4$ transition, having been calibrated using a Ramsey sequence. Defining $\tau_1$, the population difference lifetime, and $\tau_2$, the Rabi oscillation lifetime, the data is fit with the equation
\begin{eqnarray}\label{eq:rabi_fit}
\mathrm{OD} &=& A - B\exp(-dt_{mw}/\tau_1) \nonumber \\
& &+ C\exp(-dt_{mw}/\tau_2)\sin(\Omega\, dt_{mw} + \phi),
\end{eqnarray}
where $A$, $B$, $C$, $\phi$, $\tau_1$, $\tau_2$, and $\Omega$ are fitting parameters. We obtain $\tau_1=(231\pm 9) \,\mu$s and $\tau_2=(94\pm3) \, \mu$s. The Rabi oscillation lifetime is significantly shorter than the $T_2$ time obtained from the Ramsey measurement, principally due to the sensitivity of the Rabi oscillations to inhomogeneous dephasing induced by a spatially non-uniform microwave field. On the $i=4$ transition, we are sensitive to the $\pi$ component of the microwave magnetic field, and so $\Omega_{4}=2\pi \times 50.39\pm 0.05$~kHz corresponds to $B_{\pi}=3.600 \pm 0.003$~$\mu$T. We observe a strong variation in $\Omega$ across the cell (see section~\ref{sec:imaging}). The Rabi data in Fig.~\ref{fig:franzen_ramsey_rabi_inset}c was taken using a small diameter laser beam in a section of the cell with a relatively homogeneous microwave magnetic field, corresponding to a maximised $\tau_2$.

\subsection{\label{sec:further_measurements} Temperature Dependence of Relaxation Times}

\begin{figure}
\includegraphics[width=0.5\textwidth]{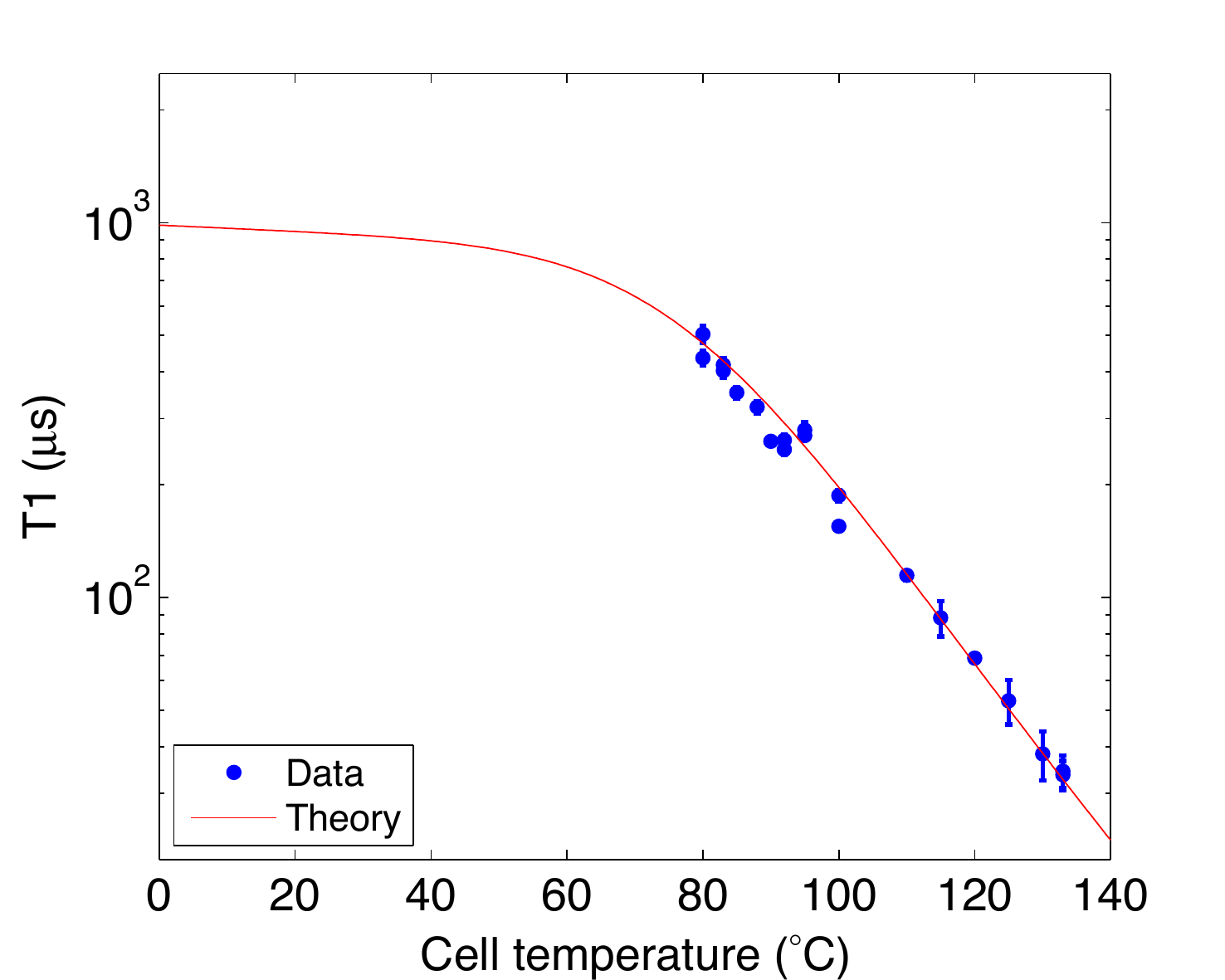}%
\caption{\label{fig:T1_T2_temperature} $T_1$ times as a function of temperature. Error bars are 95$\%$ confidence bounds from the fitting. The theory curve shows a calculation of $T_1$ using Eq.~(\ref{eq:T1_lowest_order}) with no free parameters.}
\end{figure}

Figure~\ref{fig:T1_T2_temperature} shows $T_1$ times for a range of cell temperatures, obtained using Franzen sequences measured with the photodiode. These are compared with a simple model described in~\cite{Franzen1959,Vanier1974}, which includes the effect of Rb-Rb spin exchange collisions, Rb-buffer gas collisions, atomic diffusion and atom-wall collisions. Considering only the lowest-order diffusion mode, the $T_1$ time is calculated as
\begin{equation}\label{eq:T1_lowest_order}
T_1 = [(\mu_1^2 + \nu_1^2)D + \gamma]^{-1}.
\end{equation}
Here, the diffusion coefficient is $D=D_0 P_0/P$, where $D_0$ is the diffusion coefficient at atmospheric pressure $P_0$, and $P$ is the buffer gas pressure. For a cell length $d$ and radius $R$, $\nu_1 = \pi/d$, and $\mu_1$ is defined by the first root of $J_0(\mu_1 R)=0$, where $J_0$ is the Bessel function of the first kind. The relaxation rate $\gamma = \gamma_{SE} + \gamma_\mathrm{buffer}$ accounts for relaxation due to Rb-Rb spin exchange collisions~\cite{Walter2002} at a rate $\gamma_{SE}$, and Rb-buffer gas collisions~\cite{Wagshul1994} at a rate $\gamma_\mathrm{buffer}$.
The parameters of the model are temperature-dependent; their values at $90^{\circ}$C are $\gamma_{SE}=1957\,\mathrm{s}^{-1}$, $\gamma_\mathrm{buffer}= 10\,\mathrm{s}^{-1}$, and $P=65\,\mathrm{mbar}$. For $D_0$, we use an average of the values reported in Refs~\cite{Zeng1985,Wagshul1994}, corresponding to $D_0=0.22\,\mathrm{cm}^2/\mathrm{s}$ at 90$^{\circ}$C.

At low temperatures, relaxation is governed by Rb collisions with the cell walls, with a rate proportional to the diffusion coefficient $D$. As the temperature is increased, Rb-Rb spin-exchange collisions rapidly come to dominate, due to the Rb vapor density increasing almost exponentially with temperature~\cite{Siddons2008,Nesmeyanov1963}. There is good agreement between our data and the theory, particularly at spin-exchange dominated high temperatures.

\section{\label{sec:imaging} Spatially Resolved Imaging of Relaxation Times and Microwave Field Strength}

\begin{figure*}
\includegraphics[width=1\textwidth]{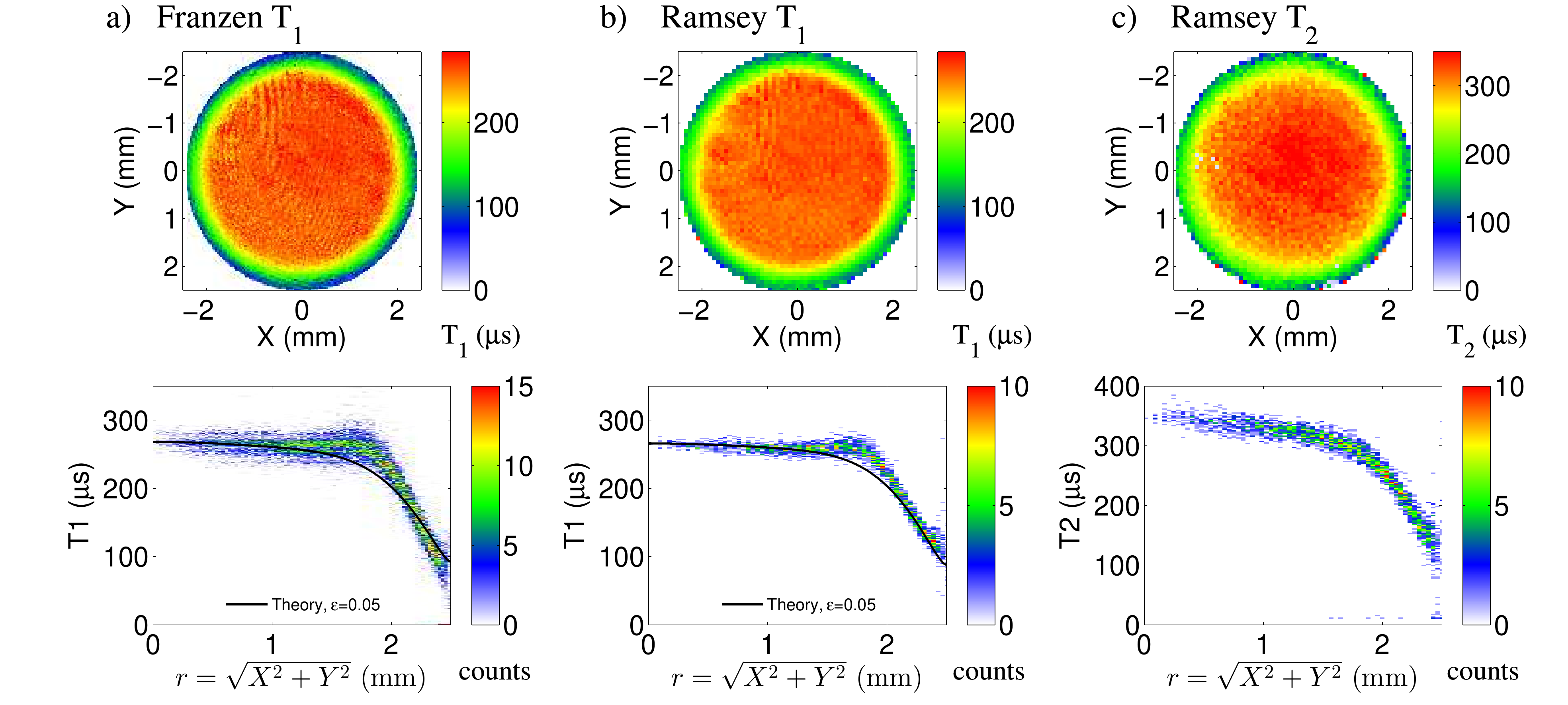}%
\caption{\label{fig:T1_T2_images} Measured $T_1$ and $T_2$ times across the cell. The top panels show a) $T_1$ times obtained from the $1/e$ decay time of a Franzen sequence (see text); b) $T_1$ times obtained from fitting a Ramsey sequence, fitting uncertainty $\pm1\%$; and c) $T_2$ times obtained from the same Ramsey sequence, fitting uncertainty $\pm4\%$. The bottom panels show radial profiles of each image in the form of a two-dimensional histogram. The radial distance from the cell center is binned into $27.5\,\mu$m wide bins for the Franzen data, and $38.8\,\mu$m wide bins for the Ramsey data. Franzen $T_1$ and Ramsey $T_1$ and $T_2$ times are binned into $0.99\,\mu$s, $1.4\,\mu$s, and $2.1\,\mu$s wide bins, respectively. The $T_1$ profiles are compared to theory as described in section~\ref{sec:modelling}. Close to the walls, there is a significant decrease in $T_1$ and $T_2$ due to Rb-wall collisions.}
\end{figure*}

We now turn our attention to measurements using the CCD camera. A single lens is used to create a 1:2 demagnified image of the cell. An ND filter is placed between the vapor cell and camera to avoid saturation of the CCD. As the camera does not have a mechanical shutter, the optical pumping pulse hits the CCD as well. The electronic shutter of the camera opens with a delay of 12~$\mu$s after the end of the pumping pulse. While some residual charges accumulated during pumping are visible on the images, they can be compensated for by taking a dark image as explained below. For the data presented in this section, the laser intensity averaged over the 5~mm cell diameter was set to $30\,\mathrm{mW}/\mathrm{cm}^2$ to obtain strong optical pumping, which ensures a large signal amplitude. During probing, on the other hand, optical pumping is undesired, and a short probe pulse duration of 2.2~$\mu$s was chosen. The strong collisional and Doppler broadening of the optical transition ensure that the transition is not strongly saturated and the number of absorbed probe photons per atom is of order unity. In an optimised setup, separate laser beams could be used to avoid compromises between optical pumping and probing performance.

Absorption imaging is a powerful technique that was perfected in experiments with ultracold atoms to obtain accurate images of atomic density distributions in a given hyperfine state \cite{Ketterle1999,Esteve2008,Streed2012}. Here we apply this technique to our vapor cell. In absorption imaging, a set of reference and dark images is usually taken in addition to the image with the atoms. This allows one to calibrate out spatial variation of the probe laser intensity and stray light \cite{Ketterle1999}. An important difference between absorption imaging of cold atoms and a hot vapor is that the presence of the atoms cannot be easily controlled in the vapor cell, i.e. the vapor is always present in the laser beam path. However, we can still modify the experimental sequence between the different images in order to be able to extract the relevant information from the observed variation in optical density $\Delta \mathrm{OD}$.

We record four images to create an image of $\Delta \mathrm{OD}$: the actual image ($I_\mathrm{image}$), taken after the entire sequence of optical pumping, microwave pulses (for Rabi and Ramsey sequences), and probe pulse; a reference image ($I_\mathrm{ref}$), taken 10~ms after every actual image, with a probe pulse, but without optical pumping or microwave pulse; a dark image for the actual image ($I_\mathrm{dark1}$), taken with a pump pulse, but no probe or microwave pulse; and a dark image for the reference image ($I_\mathrm{dark2}$), taken without any pump, probe, or microwave pulse. The two dark images are taken approximately once per day. The $\Delta \mathrm{OD}$ image is obtained by calculating
\begin{equation}\label{eq:absorption_imaging}
\Delta \mathrm{OD} = -\ln\Big[\frac{I_\mathrm{image}-I_\mathrm{dark1}}{I_\mathrm{ref}-I_\mathrm{dark2}}\Big].
\end{equation}
The absolute OD can then be determined by normalising to the unpumped value of $\mathrm{OD}=1.1$ at the cell temperature of 90$^{\circ}$C (see Figure~\ref{fig:ODvsT}). The use of reference and dark images significantly reduces our sensitivity to short and long term drifts in the imaging system and to spatial variations of the probe laser intensity. Mechanical vibrations proved to be a significant experimental challenge in achieving reliable imaging. We were required to undertake steps in order to minimise them, such as rigidising mounting components.

After taking each image, we bin the CCD pixels. This binning acts to reduce noise on the pixels and to reduce the computational intensity of the fitting process. We bin the simpler Franzen data into $3\times3$ blocks, and the Ramsey and Rabi data into $7\times7$ blocks. Taking the approximate 1:2 demagnification given by the imaging lens into account, each of these $3\times3$ ($7\times7$) pixel blocks corresponds to $35\,\mu\mathrm{m}\times35\,\mu\mathrm{m}$ ($82\,\mu\mathrm{m}\times82\,\mu\mathrm{m}$) in the cell. The spatial resolution of our imaging system is then $35\,\mu\mathrm{m}$ for Franzen data, and $82\,\mu\mathrm{m}$ for Ramsey and Rabi data. The expected size of the smallest features in the atomic vapor, on the other hand, is given by atomic diffusion through the buffer gas during the measurement sequence, typically a few hundred $\mu$m (see section~\ref{sec:imaging_relaxation} below). In the rest of this paper, we use `pixel' to refer to the $3\times3$ and $7\times7$ blocks.

\subsection{Imaging Relaxation in the Cell} \label{sec:imaging_relaxation}

Figure~\ref{fig:T1_T2_images} shows images of the $T_1$ and $T_2$ times across the cell, taken using both Franzen and Ramsey sequences. For the Ramsey sequence, the microwave input power to the cavity was 21.8~dBm, and the frequency was set slightly detuned from the $i=4$ transition.

Two different methods have been employed to obtain $T_1$ times from the Franzen and Ramsey data. Each pixel of the Ramsey data was fit using Eq.~(\ref{eq:ramsey_fit}), yielding $T_1$ and $T_2$ times with $\pm1\%$ and $\pm4\%$ fitting uncertainties, respectively. Fitting each pixel of the Franzen data in a similar way, using Eq.~(\ref{eq:franzen_fit}), yields essentially the same $T_1$ image as obtained from the Ramsey data. However, relaxation near the cell walls is not well-described by a single exponential. The model presented in section~\ref{sec:modelling}  defines $T_1$ as the $1/e$ decay time of the hyperfine population difference (Eq.~(\ref{eq:theory_T_definition})). The Franzen $T_1$ image has therefore been produced using this definition.

The bottom panels of Figure~\ref{fig:T1_T2_images} show radial profiles of the $T_1$ and $T_2$ images. There is strong agreement between the structure of the Franzen and Ramsey $T_1$ images. The relaxation rate is uniform across the centre of the cell, with both Franzen and Ramsey $T_1$ times around 265~$\mu$s. Franzen and Ramsey $T_1$ times drop away to around 80~$\mu$s and 100~$\mu$s, respectively, at the cell edge, due to the depolarisation of Rb atoms after collisions with the cell walls. The $0.34\pm0.05\,\mathrm{mm}$ half-width of this `skin' of reduced relaxation times is determined by the distance $\Delta x$ an atom diffuses during the bulk relaxation time. A simple estimate yields $\Delta x = \sqrt{D T_1} = 0.31\,\mathrm{mm}$, using the measured bulk $T_1=265\,\mu$s. More detailed modelling is described in section~\ref{sec:modelling} below. The shorter Franzen $T_1$ at the cell edge is due to the  definition of the $1/e$ time that accounts for the multimode nature of the diffusional relaxation.
The  $T_2$ relaxation, shown in the right-hand panels of Figure~\ref{fig:T1_T2_images}, also exhibits an outer `skin' of reduced relaxation times, with $T_2$ times around $130\,\mu$s at the cell edge. Unlike in the $T_1$ profiles however, the bulk $T_2$ times are not entirely flat, rising up to around $350\,\mu$s in the cell centre. 

The relaxation times obtained in the centre of the cell are larger than the values obtained using the photodiode in section~\ref{sec:PD_measurements}. Integrating over the images in Fig.~\ref{fig:T1_T2_images}a-c, we get average Franzen and Ramsey $T_1$ times of $176\,\mu$s and $221\,\mu$s, respectively, and an average Ramsey $T_2$ time of $269\,\mu$s. The Franzen $T_1$ time is more accurate, as it accounts for the multimode diffusional relaxation near the cell walls. The photodiode values lie between the central and average image values, indicating that the photodiode measurements averaged the relaxation time over some partial fraction of the cell. 

\begin{figure}
\includegraphics[width=0.5\textwidth]{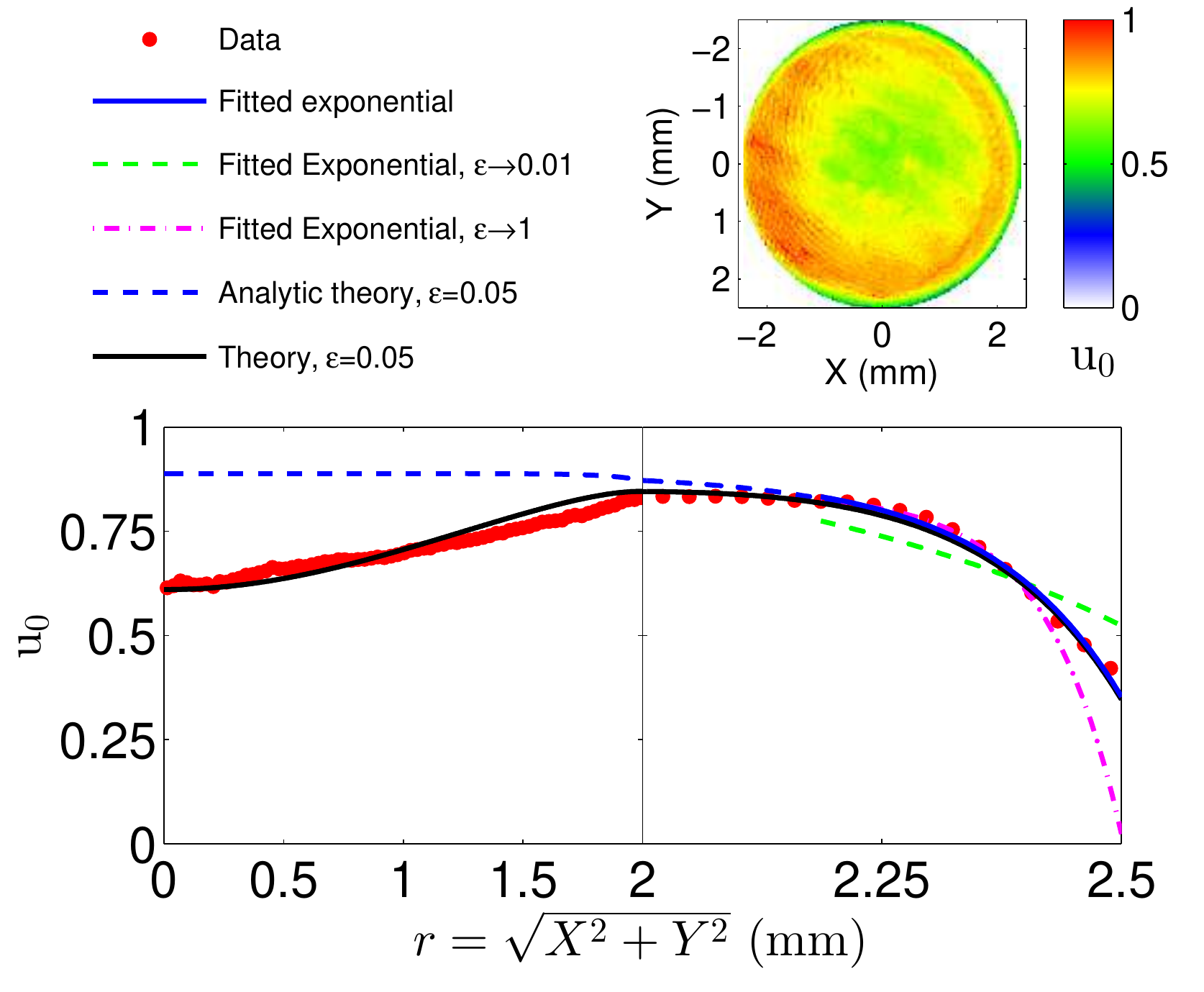}%
\caption{\label{fig:u0_franzen_inset} Image and radial profile of $u_0$, the hyperfine population difference in the optically pumped steady state, obtained from Franzen data. 
The red data points in the lower panel show the mean $u_0$ for each radial position, binned in $27.5\,\mu\mathrm{m}$ bins. The error of the mean is smaller than the symbols. 
Note the change in scaling of the bottom axis at $r=2$~mm to magnify the region near the cell wall.
The data is compared to theory as described in section~\ref{sec:modelling}. The fit of Eq.~(\ref{eq:1D_solution}) to the data near the cell wall is shown in solid blue. The `theory' and `analytic theory' curves respectively model $u_0$ with (Eq.~(\ref{eq:u0_prime})) and without (Eq.~(\ref{eq:u0})) the inclusion of the central dip in optical pumping efficiency, which was caused by a Rb deposit on the front cell window. }
\end{figure}

In addition to the relaxation times, the absorption images also provide information about the optical pumping efficiency.
We define the hyperfine population difference between the  $F=1$ and $F=2$ states as
\begin{equation}\label{eq:u_definition}
u \equiv 1 - \frac{n_2}{5/8} ,
\end{equation}
where $n_2$ is the fraction of atoms in $F=2$. With this definition, $u=0$ represents the unpumped equilibrium state where all $m_F$ states are equally populated, and $u=1$ corresponds to perfect optical pumping where the $F=2$ state is empty.
The $B$ fitting parameter for Franzen data (see Eq.~(\ref{eq:franzen_fit})) describes the amount the OD has changed through optical pumping. Normalising by the unpumped $\mathrm{OD}=1.1$ we obtain the hyperfine population difference in the optically pumped steady state, $u_0=B/1.1$.
Figure~\ref{fig:u0_franzen_inset} shows the image and radial profile of $u_0$ obtained in this way.
We observe a reduced pumping efficiency close to the cell edge because of atom-wall collisions. In addition, there is a broad dip in  pumping efficiency in the centre of the cell. This is due to a deposit of Rb that had developed on the front cell wall, partially blocking the pumping light. The deposit was present when taking all of the imaging data. The robustness of our $T_1$, $T_2$, and microwave magnetic field measurements is highlighted by the lack of correlation between the image of $u_0$ in Figure~\ref{fig:u0_franzen_inset}, and the images presented in Figures~\ref{fig:T1_T2_images}~and~\ref{fig:mw_field_images}.

\subsection{\label{sec:modelling} Modelling Relaxation in the Cell}

We now describe a model for the hyperfine population relaxation in the cell and compare it with our imaging data.
We begin by analyzing the optically pumped steady state in Fig.~\ref{fig:u0_franzen_inset}. Using a simple 1D model based on Ref.~\cite{Grafstroem1996b}, we determine the probability that a Rb-wall collision destroys the hyperfine polarisation.
We then use this probability in a 2D model valid throughout the entire cell to describe the observed $T_1$ relaxation.

\subsubsection{\label{sec:1D_model} Depolarisation Probability of Rb-Wall Collisions}

In Ref.~\cite{Grafstroem1996b}, Grafstr{\"o}m and Suter used evanescent-wave spectroscopy to study optical pumping of Na vapor near a glass wall. Using a simple model, they related the atomic $\langle m_F \rangle$-polarisation at the wall to the depolarisation probability of atom-wall collisions. We adapt their model to our case of hyperfine population relaxation between states of different $F$ in Rb collisions with Si walls.

Close to the cell walls, the evolution of the hyperfine population difference $u$ can be described by a 1D diffusion equation
\begin{equation}\label{eq:diffusion_equation_1D}
\frac{\partial}{\partial t}{u}(r,t) = D \frac{\partial^2 u}{\partial r^2} - (\Gamma +\Gamma_p) u(r,t)+\Gamma_p.
\end{equation}
The first term on the right-hand-side describes diffusion of Rb atoms in the buffer gas. The second term describes relaxation at a rate $\Gamma+\Gamma_p$, where the bulk relaxation rate $\Gamma = \gamma_{SE} + \gamma_\mathrm{buffer}+\gamma_z$ includes the effect of Rb-Rb spin exchange collisions ($\gamma_{SE}$) and Rb-buffer gas collisions ($\gamma_\mathrm{buffer}$). Relaxation due to collisions with the front and back cell windows varies only slightly with $r$, and so we include it as a constant rate $\gamma_z$. The optical pumping rate $\Gamma_p$ drives both relaxation in the second term of Eq.~(\ref{eq:diffusion_equation_1D}) and optical pumping in the third term.
The steady-state solution to Eq.~(\ref{eq:diffusion_equation_1D}) is
\begin{equation}\label{eq:1D_solution}
u_0(r) = u_{\infty} - (u_{\infty}-u_R)\exp[\mu (r-R)],
\end{equation}
where $u_{\infty} \equiv \frac{\Gamma_p}{\Gamma + \Gamma_p}$ is the population difference far from the walls, $R$ is the cell radius, $u_R \equiv u_0(R)$ is the population difference at the wall, and $\mu \equiv \sqrt{\frac{\Gamma+\Gamma_p}{D}}$.
Wall collisions produce a skin of reduced optical pumping near the cell edge, with the skin thickness given by $\mu^{-1}$.
The 1D model provides a good description of the behavior near the wall for $|r-R| \ll R$ and $\mu R \gg 1$, which is satisfied in our experiment.

From the behavior of $u_0(r)$ near the cell wall, it is possible to determine the probability $\epsilon$ that a Rb-wall collision destroys the atomic hyperfine polarisation \cite{Grafstroem1996b}. Very close to the wall, on average half of the atoms have just collided with the wall, and half are arriving from a distance $L=\tfrac{2}{3}\lambda$ into the cell bulk, where $\lambda = 3.5\,\mu$m is the Rb mean free path in the buffer gas. Atoms from the bulk carry an average polarisation $u(R-L)$, which is reduced to $(1-\epsilon) u(R-L)$ after the collision. Thus, $u(R) \simeq \tfrac{1}{2}(2-\epsilon) u(R-L)$. Applying these considerations to Eq.~(\ref{eq:1D_solution}) and exploiting that $\mu L \ll 1$, we obtain
\begin{equation}\label{eq:epsilon}
\epsilon = \frac{2\mu L (u_{\infty}-u_R)}{u_R + \mu L (u_{\infty}-u_R)}.
\end{equation}
Figure~\ref{fig:u0_franzen_inset} shows a fit of Eq.~(\ref{eq:1D_solution}) to the measured $u_0(r)$ profile of the Franzen data (blue solid line). We only fit to the data near the cell wall ($r\ge2.15$~mm), where the 1D approximation is valid and the optical pumping rate is approximately constant. The fit parameters are $\mu = (7\pm1)\times10^3 \,\mathrm{m}^{-1}$, $u_R = 0.35\pm0.04$, and $u_{\infty} = 0.89\pm0.03$. Using these values in Eq.~(\ref{eq:epsilon}), we obtain a depolarisation probability of $\epsilon=0.05\pm0.01$. When we analyse the initial state of the Ramsey data in a similar way (not shown), we obtain $\epsilon=0.046\pm0.007$, consistent with the Franzen data.
For comparison, Fig.~\ref{fig:u0_franzen_inset} shows fits to the data where $\epsilon$ was constrained to $\epsilon=1$ (purple) and $\epsilon = 0.01$ (green), respectively. Both values are inconsistent with our data.

The value of $\epsilon=0.05$ obtained from our data is surprisingly small. It implies that the atomic hyperfine population can survive of order $\epsilon^{-1} \approx 20$ collisions with the Si wall. Previous experiments with Na and Cs atoms near glass walls have reported $\epsilon=0.5$ \cite{Grafstroem1996b,DeFreitas2002}. Our experiment differs not only in the measurement technique, the atomic species, and the wall material, but also in that we study relaxation between hyperfine states $F=2$ and $F=1$, while the previous experiment \cite{Grafstroem1996b} studied the relaxation of $\langle m_F \rangle$-polarisation within one hyperfine state. A systematic error in our measurement would arise if the images are clipped close to the cell wall, so that the actual location of the wall is at $r>2.5$~mm. To make our data consistent with $\epsilon = 1$, the location of the wall would have to be shifted by $>63\,\mu$m (more than two datapoints in Fig.~\ref{fig:u0_franzen_inset}), which is not very likely given the spatial resolution of our imaging system. Moreover, we point out that the surface properties of the interior cell walls are not precisely known. A layer of adsorbed Rb atoms or other residues on the Si walls could modify the collisional properties. A systematic study of these effects would require a dedicated setup and is beyond the scope of the present work. However, our measurements show that absorption imaging is a powerful tool for the investigation of atom-wall collisions. The high spatial resolution opens up many intriguing possibilities such as laterally patterning the surface to modulate the collisional properties.

\subsubsection{\label{sec:2D_model} $T_1$ Relaxation: 2D Model}

We now model $T_1$ relaxation in the Franzen sequence, considering the entire circular aperture of our cell. The diffusion equation for circular symmetry reads
\begin{equation}\label{eq:diffusion_equation_2D}
\frac{\partial}{\partial t}{u}(r,t) = D \frac{1}{r}\frac{\partial}{\partial r}\left( r \frac{\partial u(r,t)}{\partial r} \right) - (\Gamma +\Gamma_p) u(r,t)+\Gamma_p.
\end{equation}
From the above considerations on diffusion and atom-wall collisions, we can derive the boundary condition
\begin{equation}\label{eq:boundary_conditions}
\frac{\partial u}{\partial r}\Big|_{r=R} + \frac{\epsilon/2}{(1-\epsilon/2)L}u(R) = 0,
\end{equation}
which reproduces Eq.~(\ref{eq:epsilon}) when applied to Eq.~(\ref{eq:1D_solution}).
The initial condition for modeling the Franzen sequence is given by the optically pumped steady state solution of Eq.~(\ref{eq:diffusion_equation_2D}) subject to the boundary condition Eq.~(\ref{eq:boundary_conditions}),
\begin{equation}\label{eq:u0}
u_0(r) = u_{\infty} \Big(1-\frac{I_0(\mu r)}{I_0(\mu R)+I_1(\mu R)\,(2/\epsilon-1)\mu L} \Big),
\end{equation}
where $I_0$ and $I_1$ are modified Bessel functions of the first kind, and $u_{\infty}$ and $\mu$ are defined as in the previous section. In the following, we take $\epsilon=0.05$ as a fixed parameter determined as described above.

Figure~\ref{fig:u0_franzen_inset} shows $u_0(r)$ given by Eq.~(\ref{eq:u0}) for the same parameters as in the previous section (blue dotted line). While the solution is indistinguishable from the 1D model close to the wall and matches the data well in this region, there is a discrepancy in the cell center ($r<2$~mm). This is because we have so far assumed a spatially homogeneous optical pumping rate $\Gamma_p$, which was not the case in the experiment. To model $T_1$ relaxation, we can simply take the measured profile in Fig.~\ref{fig:u0_franzen_inset} as the initial condition for the dynamics described by Eq.~(\ref{eq:diffusion_equation_2D}). It can be phenomenologically described by the function
\begin{eqnarray}\label{eq:u0_prime}
u^\prime_0(r) = u_0(r) - \frac{k_0}{2} \left[ \cos \left(\pi \frac{r}{R}\right)+1 \right].
\end{eqnarray}
The additional term has been chosen such that it does not affect the boundary condition Eq.~(\ref{eq:boundary_conditions}) and is thus consistent with the same value of $\epsilon$ as $u_0(r)$. The factor $k_0$ describes the reduced pumping efficiency in the cell center. Our data is well described by $u^\prime(r)$ using $k_0=0.28$ (black solid line in Fig.~\ref{fig:u0_franzen_inset}).

\begin{figure*}
\includegraphics[width=1\textwidth]{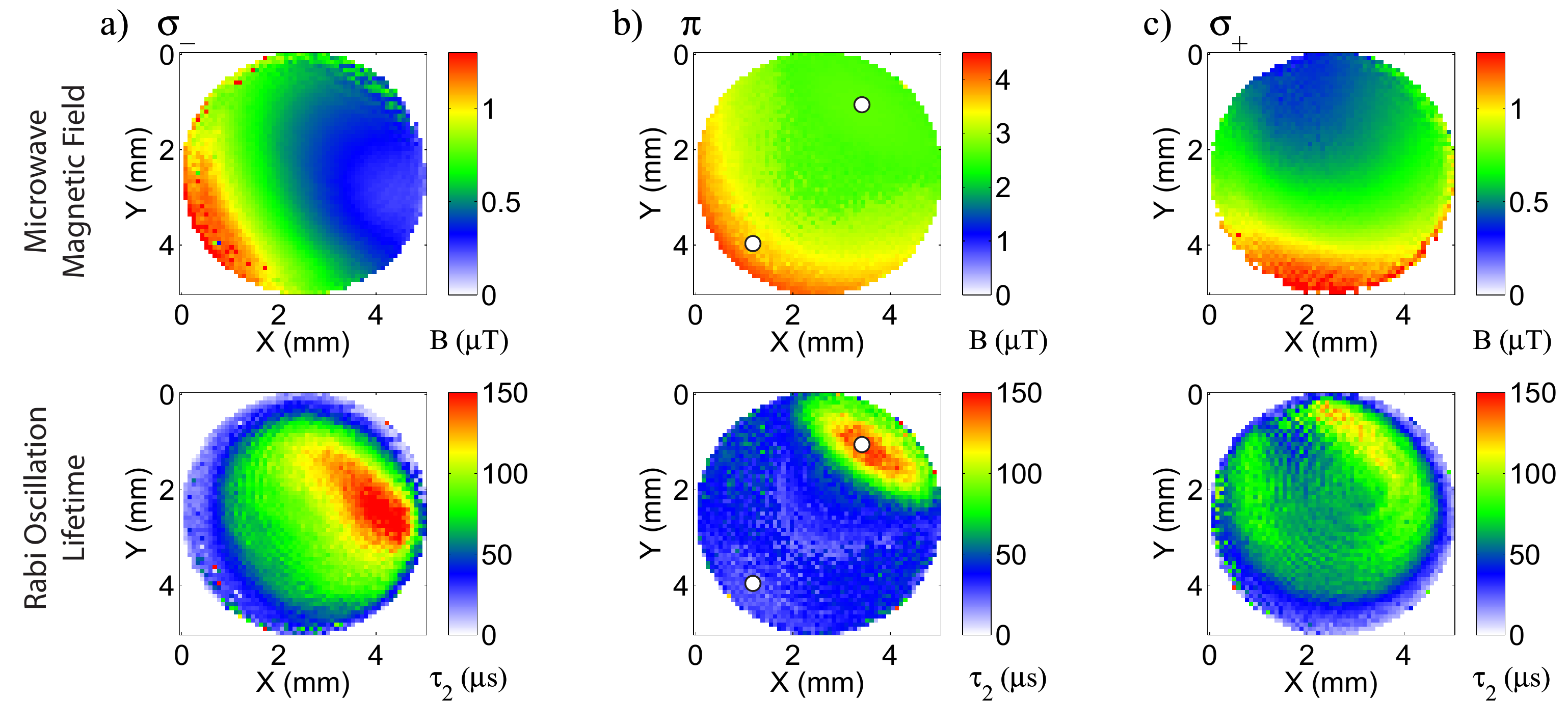}%
\caption{\label{fig:mw_field_images} Top: Rabi sequences have been used to obtain images of the a) $\sigma_-$, b) $\pi$, and c) $\sigma_+$ components of the microwave magnetic field. Bottom: Images of the corresponding Rabi oscillation lifetimes, $\tau_2$. The white dots on the $\pi$ images show the approximate locations of the pixels examined in Figure~\ref{fig:rabi_pi_pixel_analysis}.}
\end{figure*}

We model relaxation in the dark by setting $\Gamma_p=0$ at $t\geq0$ and numerically solving Eq.~(\ref{eq:diffusion_equation_2D}) with the initial condition Eq.~(\ref{eq:u0_prime}) and the boundary condition Eq.~(\ref{eq:boundary_conditions}). At each radial position, we define $T_1$ as the time taken for $u$ to decay to $1/e$ of its initial value:
\begin{equation}\label{eq:theory_T_definition}
u(r,T_1)=\frac{1}{e} u(r,0).
\end{equation}
In the limit where the temporal decay of $u$ can be described by a single exponential, this definition is identical to that used in the fits of section~\ref{sec:PD_measurements}.
The simulated and measured $T_1$ profiles are compared in the bottom panels of Fig.~\ref{fig:T1_T2_images}.
We set $\Gamma=3900\,\mathrm{s}^{-1}$ in order to match the theory curves with the observed $T_1$ values in the centre of the cell. The central dip in optical pumping efficiency results in $T_1>\Gamma^{-1}$ in the cell center due to the diffusive influx of atoms from neighbouring regions with higher optical pumping, partially offsetting relaxation. The agreement of our model with the data is reasonable. In particular, the width of the skin of reduced $T_1$ times at the cell edge is reproduced well. However, the transition from the cell bulk to the cell edge is sharper in the data than in the model.

\subsection{Imaging the Microwave Field}

Figure~\ref{fig:mw_field_images} shows images of the $\sigma_-$, $\pi$, and $\sigma_+$ components of the microwave magnetic field, obtained using Rabi measurements on transitions $i=1$, 4, and 7, respectively. The bottom panels show the corresponding decay times of the Rabi oscillations ($\tau_2$). The microwave frequency was calibrated using Ramsey sequences, and tuned exactly to resonance for each transition. The microwave power at the input to the cavity was 26.8~dBm. Each pixel was fit using Eq.~(\ref{eq:rabi_fit}), and the microwave magnetic field strength was then calculated using Eqs.~(\ref{eq:mw_Bfield}).

The principal component of the cavity microwave magnetic field is the $\pi$ component, with a strength more than 3 times that of the $\sigma$ components. The dominance of the $\pi$ component follows from the cavity design~\cite{Mileti1992,Stefanucci2012}. The presence of the $\sigma$ components is not unexpected, as we are using a much smaller vapor cell than the one the cavity was designed for, and both the cavity tuning and field geometry are strongly dependent upon the dielectric filling provided by the glass and silicon cell walls. This non-optimal dielectric charging of the cavity is likely, in addition, to be the reason for the relatively high inhomogeneity measured for the microwave field. Such inhomogeneities are undesirable for most applications of the cavity, but here they aid in the demonstration of our imaging technique and its capabilities. It is also possible that the inhomogeneities are caused by some microwave field radiated directly from the loop coupling the microwave into the cavity: while the 6.8~GHz microwave frequency is below cutoff with respect to the outer cylinder of the cavity, in these images we are using an input microwave power several orders of magnitude above the -30 to -10 dBm typically used for clock applications.

The lifetime, $\tau_2$, of the Rabi oscillations is significantly shorter than the $T_2$ time, principally due to inhomogeneities in the microwave magnetic field~\cite{Arditi1964}. This can be seen in Figure~\ref{fig:mw_field_images}, where the $\tau_2$ time is inversely correlated with the magnitude of the microwave magnetic field inhomogeneity, which in turn is linked to the field strength. We see that this effect is strongest for oscillations on the $i=4$ transition, corresponding to the $\pi$ component of the field. The $\tau_2$ values on this transition are only $20$-$40\,\mu$s across much of the cell. In the field minimum of each transition, where inhomogeneities are smallest, $\tau_2$ is around $150\,\mu$s.

As a higher field strength also drives faster oscillations, the number of visible oscillations is a measure of the quality of the coherent driving. We find that this number remains roughly constant across most of the images, with 1-2 oscillations visible over the $\tau_2$ time. The high $\tau_2$ region in the upper right of the $\pi$ image (Figure~\ref{fig:mw_field_images}.b), with $\tau_2$ values around 150~$\mu$s, is an exception: In this region, more than 5 oscillations are visible. It is not clear why there is such a local increase in the number of visible oscillations, as this is not seen in the high $\tau_2$ regions on the $\sigma$ transitions.

\begin{figure}
\includegraphics[width=0.5\textwidth]{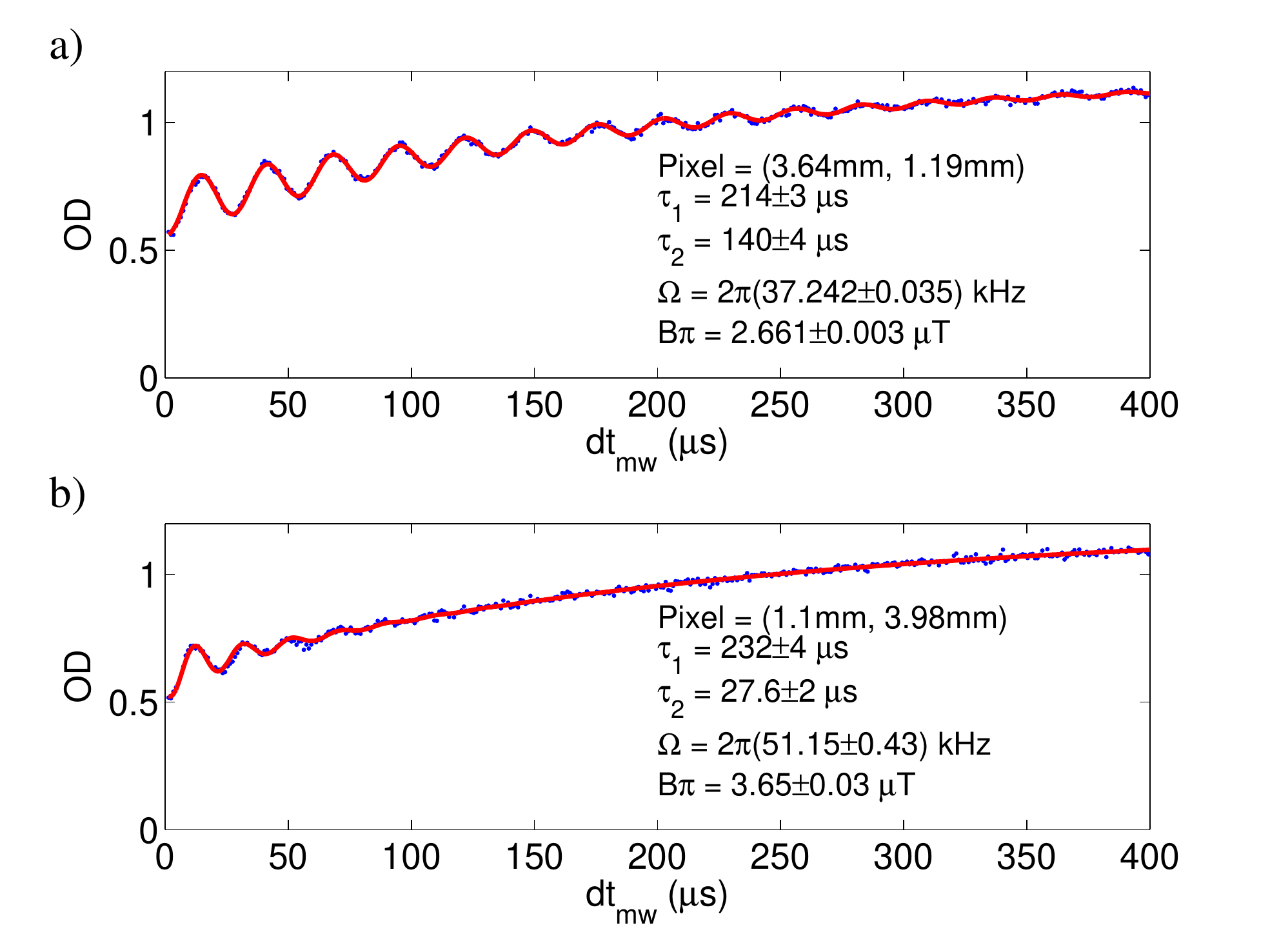}%
\caption{\label{fig:rabi_pi_pixel_analysis}
Representative pixels of the $\pi$ images in Figure~\ref{fig:mw_field_images}. Fitted data is shown for pixels in a) the high $\tau_2$ region ($x=3.64\,\mathrm{mm}$, $y=1.19\,\mathrm{mm}$) and b) the low $\tau_2$ region ($x=1.10\,\mathrm{mm}$, $y=3.98\,\mathrm{mm}$). Atoms in the high $\tau_2$ region perform an unusually large number of Rabi oscillations.}
\end{figure}

Figure~\ref{fig:rabi_pi_pixel_analysis} shows examples of Rabi oscillations for two representative pixels from the $\pi$ image (marked by white dots in Figure~\ref{fig:mw_field_images}). The top panel shows a pixel from the high $\tau_2$ region, ($x=3.64\,\mathrm{mm}$, $y=1.19\,\mathrm{mm}$), while the bottom panel shows a pixel with low $\tau_2$, ($x=1.10\,\mathrm{mm}$, $y=3.98\,\mathrm{mm}$). Atoms in the high $\tau_2$ region can be seen to undergo many more Rabi oscillations than atoms in the rest of the cell.

The images show that different hyperfine transitions can have quite spatially different regions of optimal $\tau_2$, depending on the geometry of the applied microwave field. The strong spatial variation in $\tau_2$ highlights the importance of our technique for cell and cavity characterisation, in particular for high precision devices such as vapor cell atomic clocks.

\section{\label{sec:outlook} Conclusions and Outlook}

We have used time-domain spatially resolved optical and microwave measurements to image atomic relaxation and the polarisation-resolved microwave magnetic field strength in a microfabricated Rb vapor cell placed inside a microwave cavity. The population relaxation times were measured to be approximately uniform across the cell centre, with a value at $90^{\circ}$C of $T_1=265\,\mu$s, whilst coherence times in the cell centre peaked at around $T_2=350\,\mu$s. Depolarising collisions between Rb atoms and the cell walls resulted in $T_1$ and $T_2$ times around $80\,\mu$s and $130\,\mu$s near the cell walls, respectively. Diffusion of these atoms lowered relaxation times within $0.7\,$mm of the cell wall.

The relaxation times at the cell edge provide spatially resolved information on the interactions of Rb atoms with the silicon cell walls. Our data suggest that Rb-wall collisions are not completely depolarising, agreeing with previous work. This aspect of our technique could be particularly useful in the characterisation of wall coatings in coated cells.

Images of the cavity microwave magnetic field show significant spatial inhomogeneity in each of its three vector components, $\sigma_-$, $\pi$, and $\sigma_+$, due to perturbations to the cavity introduced by the dielectric cell material. For each vector component, we can identify the resulting region maximising the number of Rabi oscillations, and hence the region of optimal coherent manipulation.

Our measurement technique is fast, simple, and produces high resolution images for vapor cell and microwave-device characterisation. It is of particular interest for characterising cells in miniaturised atomic clocks~\cite{Knappe2007} and sensing applications~\cite{Schwindt2007,Donley2009,Bohi2012}. It is also of interest for characterising the cell and cavity properties in larger and high-performance vapor cell atomic clocks~\cite{Micalizio2012,Affolderbach2006,Bandi2011}.




\begin{acknowledgments}
This work was supported by the Swiss National Science Foundation (SNFS) and the European Space Agency (ESA). We thank Y. P\'{e}tremand for filling the cell, and R. Schmied, J. Kitching, L. Weller and I. Hughes for helpful discussions.
\end{acknowledgments}

\bibliography{bibliography_cell_characterisation}

\end{document}